\documentclass[useAMS,usenatbib]{mn2e}
\usepackage{graphics}
\usepackage[dvips]{epsfig}
\usepackage{lscape}
\usepackage{multirow}
\usepackage{natbib}
\usepackage{aas_macros}
\usepackage{amsmath}
\usepackage{epsfig}
\usepackage[table]{xcolor}
\def\grb{GRB\,080210}

\title[ULTRASPEC fast imaging of \grb{}]{Probing GRB environments with time variability: ULTRASPEC fast imaging of \grb{}
\thanks{Based on observations collected with the ULTRASPEC visitor instrument built by a consortium from the University of Sheffield, Warwick, the UK Astronomy Technology Centre and ESO, mounted at the ESO/3.6-m telescope on La Silla, Chile, and on target-of-opportunity observations collected in service mode under program ID 080.D-0526, P.I. Vreeswijk, with the FOcal Reducer/low dispersion Spectrograph 2 (FORS2) installed at the Cassegrain focus of the Very Large Telescope (VLT), Unit 1, Antu, operated by the European Southern Observatory (ESO) on Cerro Paranal in Chile. For further information or questions on the content of the paper, please e-mail to annalisa@raunvis.hi.is.}}

\author[De Cia et al.]
{A. De Cia$^{1}$, P. Jakobsson$^{1}$, G. Bj\"{o}rnsson$^{1}$, P. M. Vreeswijk$^{1,2}$, V. S. Dhillon$^{3}$, \and T. R. Marsh$^{4}$, R. Chapman$^{1,5}$, J. P. U. Fynbo$^{2}$, C. Ledoux$^{6}$, S. P. Littlefair$^{3}$, \and D. Malesani$^{2}$, S. Schulze$^{1}$, A. Smette$^{6}$, T. Zafar$^{2}$ and  E. H. Gudmundsson$^{1}$.\\
$^{1}$ Centre for Astrophysics and Cosmology, Science Institute, University of Iceland, Dunhaga 5, IS-107 Reykjavik, Iceland\\ 
$^{2}$ Dark Cosmology Centre, Niels Bohr Institute, University of Copenhagen, 2100 Copenhagen \O, Denmark\\ 
$^{3}$ Department of Physics and Astronomy, University of Sheffield, Sheffield S3 7RH, UK\\ 
$^{4}$ Department of Physics, University of Warwick, Coventry CV4 7AL, UK\\ 
$^{5}$ Centre for Astrophysics Research, University of Hertfordshire, College Lane, Hatfield AL10 9AB, UK\\ 
$^{6}$ European Southern Observatory, Alonso de C\'ordova 3107, Vitacura, Casilla 19001, Santiago 19, Chile\\
}

\begin{document}
\date{Accepted 2010 Mm gg. Received 2010 Mm gg; in original form 2010 Mm gg}
\pagerange{\pageref{firstpage}--\pageref{lastpage}} \pubyear{2010}
\maketitle

\label{firstpage}

\begin{abstract}
 We present high time resolution (1.09 s) photometry of \grb{} obtained with ULTRASPEC mounted on the ESO/3.6-m telescope, starting 68.22 min after the burst and lasting for 26.45 min. The light curve is smooth on both short (down to 2.18 s) and long time scales, confirmed by a featureless power spectrum. On top of the fireball power-law decay, bumps and wiggles at different time scales can, in principle, be produced by density fluctuations in the circumburst medium, substructures in the jet or by refreshed shocks. Comparing our constraints with variability limits derived from kinematic arguments, we exclude under-density fluctuations producing flux dips larger than 1 per cent with time scales $\Delta t > 9.2$ min (2 per cent on $\Delta t > 2.3$ min for many fluctuating regions). In addition, we study the afterglow VLT/FORS2 spectrum, the optical-to-X-ray spectral energy distribution (SED) and the time decay. The SED is best fit with a broken power law with slopes $\beta_{\mathrm{opt}}=0.71\pm0.01$ and $\beta_{X}=1.59\pm0.07$, in disagreement with the fireball model, suggesting a non-standard afterglow for \grb{}. We find $A_V=0.18\pm0.03$ mag optical extinction due to SMC-like dust and an excess X-ray absorption of log $(N_{\mathrm{H}}/$cm$^{-2})=21.58^{+0.18}_{-0.26}$ assuming Solar abundances. The spectral analysis reveals a damped Ly$\alpha$ absorber (log $(N_{\mathrm{H\,{\sc I}}}/$cm$^{-2})=21.90\pm0.10$) with a low metallicity ([X/H$]=-1.21\pm0.16$), likely associated with the interstellar medium of the GRB host galaxy ($z=2.641$).\\ 
\end{abstract}


\section{Introduction}

\begin{center}
   \begin{table*}
   \begin{tabular}{ l l c c c r r }
\hline \hline
\rule[-0.2cm]{0mm}{0.6cm}
Instrument & Grism & Start time & Exposure time & $\delta t^a$ & Coverage & FWHM\\
\rule[-0.2cm]{0mm}{0.6cm}
 &  & (UT hh:mm:ss) &  (s) & (min) & \multicolumn{1}{c}{(\AA{})} & \multicolumn{1}{c}{(\AA{})}  \\
\hline

VLT/FORS2  & ---                    & 08:26:25  & 10 & 36.4 & $R$ band & \\

VLT/FORS2 & 300V                & 08:32:10 & 600 & 47.1 & 3500--9600 & 13.3 \\ 

VLT/FORS2 & 600z+OG590   & 08:43:13 & 600 & 58.1&  8000--9000 & 6.4 \\ 

VLT/FORS2 & 1400V              & 08:54:37 & 600 & 69.5 & 4600--5900 & 2.5  \\ 

ULTRASPEC & ---                  & 08:58:18 & $1.09\times1455$ & 81.4 & $V$ band & \\ 

VLT/FORS2 & 1200R+GG435 & 09:05:47 & 600 & 80.7& 6000--7000  & 3.0 \\  

VLT/FORS2 & 300V                & 09:17:11 & 600 & 92.1 & 3500--9600  & 13.3 \\ 

VLT/FORS2  & ---                    & 09:30:53  & 45 & 101.2 & $R$ band & \\

\hline \hline
 \end{tabular} 
\caption{\grb{} observation log on date 2008 February 10. $^a$ $\delta t$ is the mid-exposure time after the BAT trigger (07:50:06 UT).}
\label{tab1} 
  \end{table*}  
\end{center}

Long ($>2$ s) soft gamma ray bursts (GRBs) are the most powerful explosions known in the Universe. After the discovery of GRB optical \citep{VanParadijs97} and X-ray \citep{Costa97} afterglows in 1997, we have learned that they mainly occur in distant galaxies and their connection to core-collapse supernovae is now widely accepted \citep[for a review see][]{Woosley06}. The diversity amongst individual GRB events, as well as the difficulty in observing such transient sources, challenges theoretical models to explain them, the fireball model providing the best overall agreement \citep[e.g.,][]{Rees92,Meszaros93,Piran99}. In this scenario, the GRB afterglow originates from the synchrotron radiation produced by the interaction between the ultra-relativistic ejecta (jet) and the surrounding interstellar medium.\\

Although GRBs can be extremely variable in their prompt phase and X-ray afterglow flares are commonly observed during the first minutes after the burst, the late-time afterglow in general shows a fairly smooth power-law behaviour at different phases \citep{Zhang06}, from the X-rays to the optical, IR and radio wavelengths. Environmental effects and intrinsic discontinuities can introduce afterglow variability on different time scales, possibly due to \textit{(i)} ambient density fluctuations \citep{Wang00}, \textit{(ii)} substructures in the jets \citep*[patchy-jet;][]{Meszaros98}, \textit{(iii)} inhomogeneities on the emitting surface \citep[patchy-shell;][]{Kumar00}, \textit{(iv)} refreshed shocks \citep{Rees98,Sari00} or \textit{(v)} late-time central engine activity \citep{Rees00}.\\ 

\textit{(i)} Density fluctuations can arise from interstellar turbulence or can be generated, before the GRB event, by a variable wind from the progenitor star \citep[see e.g.,][]{VanMarle05}. Linear density fluctuations with $dn/n<1$ on a length scale of 1--$10^3$ AU could induce fluctuations in the afterglow light curve with a fractional amplitude of up to $\sim30$ per cent over time scales of tens of minutes in the optical \citep{Wang00}. \textit{(ii)} Substructures in the jet can form if the bulk Lorentz factor depends on the angle inside the jet \citep{Meszaros98}. As the emitting region evolves through this patchy-jet, the flux varies in intensity. \textit{(iii)} Angular inhomogeneity of the relativistic ejecta can separate the emitting surface into different causally disconnected regions \citep[patchy-shell;][]{Kumar00,Nakar04}. These wiggles evolve within the emitting surface, which is expanding in time with the blast-wave deceleration, causing variability in the radiation. \textit{(iv)} If the ejecta have a range of bulk Lorentz factors, the slower shells will catch up with the leading blast wave, once the fireball has been decelerated by the external medium. The refreshed shocks will boost the luminosity of the afterglow \citep{Rees98}. \textit{(v)} The GRB engine could contribute to the variability at late times: as debris accretes onto the black hole in the period following the burst, its extended activity could heat the environment or produce new outflows, giving rise to a detectable component of emission which, like any accretion-powered source, would be variable \citep{Rees00}. Thus, the detection (and even the non-detection) of time variability within several minutes to a couple of hours after the burst can provide important constraints on the different proposed scenarios, and therefore on the physics of the evolution of the fireball.\\
 
 Temporal variations in GRB afterglow light curves were first observed, on time scale as short as $\sim1$ hour, in GRB\,011211{} \citep{Holland02,Jakobsson04}, induced either by inhomogeneities in the medium surrounding the GRB, or by a patchy jet. \citet{Lazzati03} have argued that the deviations in the afterglow of GRB\,021004{} are due to the interaction of the GRB fireball, or jet, with density enhancements in the ambient medium. However, time-resolved polarimetry of the same burst suggested that the variations were produced by a refreshed shock \citep*{Bjornsson04}. \citet*{Granot03} have also interpreted the variations seen in the light curve of GRB\,030329 as due to refreshed shocks.\\

Several limitations challenge the detection of late-time variability in GRB light curves. First, the amplitude of most of the fluctuations that can possibly be expected decays with time. Moreover, different processes (e.g., density fluctuations, patchy shell or refreshed shocks) will physically constrain the variability over only certain time scales at one observation time \citep*{Ioka05}.  In particular, the fastest variability is the hardest to detect, partly because it is intrinsically weaker, but also because the readout noise and dead time of classical CCDs usually limits the time resolution of the observation itself. High-speed photometry can now be achieved thanks to the fast read-out with zero-noise of the frame-transfer electron-multiplying CCDs (EMCCDs). The ULTRASPEC camera \citep{Dhillon07} adopts such a CCD to amplify the signal, rendering the read-out noise negligible. In addition, the frame transfer architecture allows the EMCCD to read out a completed exposure whilst the next exposure is being obtained, virtually eliminating the dead time between exposures.\\
 
 We observed \grb{} with ULTRASPEC mounted on the ESO 3.6-m telescope in La Silla, allowing 1.09 s time resolution imaging. The highest speed photometry obtained for a GRB afterglow so far is the TORTORA  observations of the extremely bright ``naked eye'' GRB 080319B \citep{Greco09,Beskin10}, with a 0.3 s time bin, from 10 to 100 s after the burst trigger. However, the ULTRASPEC observations of \grb{}, presented in this paper, probes the afterglow phase, providing the lowest $\Delta t / t$ so far. This opens a new window on the fast-variability study of the afterglow itself. Comparing the variability limits given by \citet{Ioka05} with the ULTRASPEC observations, we can constrain the properties of the circumburst medium and the shock structure.\\ 
 
In addition, we investigate the \grb{} host galaxy environment in another way, through ESO-VLT/FORS low- and medium-resolution spectroscopy, as well as optical-to-X-ray spectral energy distribution (SED) modelling. Ly$\alpha$ and metal absorption systems, often observed in GRB lines of sight, can be used to derive physical properties of the absorbing gas clouds, such as kinematics, densities and metallicities \citep{Vreeswijk06,Ledoux09}. On the other hand, the SED provides information on both the dust inside the host galaxy and the spectral properties of the GRB afterglow itself \citep[e.g.,][]{Starling07}. Overall, interpreting the combined optical and X-ray spectra and light curves within the context of the fireball model can provide a probe of the blast-wave physics, as well as the GRB environment \citep[Sari, Piran \& Narayan 1998;][]{Zhang06}.\\

The paper is organized as follows: observations and data reduction are presented in Sec. 2, while their analysis is reported in Sec. 3: first the ULTRASPEC light curve, then the optical spectroscopy, the SED modelling and finally the optical and X-ray afterglow temporal decay. We discuss our results in Sec. 4 and summarize them in the last section. Throughout the paper we use the convention $F_\nu \left(t\right) \propto t^{-\alpha}\nu^{-\beta}$ for the flux density, where $\alpha$ is the temporal slope and $\beta$ is the spectral slope. Hereafter we assume a standard $\Lambda$CDM cosmology with $H_0$ = 70.4 km s$^{-1}$ Mpc$^{-1}$, $\Omega_M$ = 0.27 and $\Omega_{\Lambda}$ = 0.73 \citep{Jarosik10}. \\

\begin{figure}
 \centering
 \includegraphics[width=85mm,angle=0]{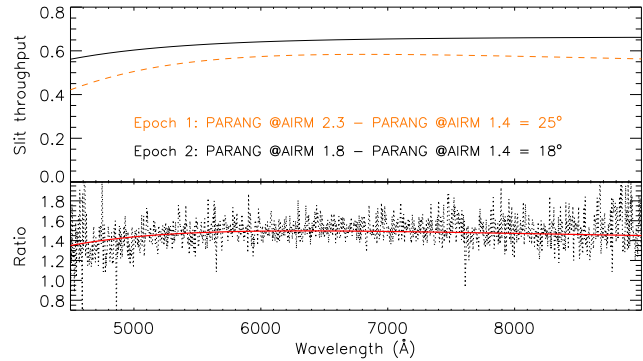}
\caption{Estimated slit throughputs (top panel) of the two 300V spectra due to the wavelength dependence of the PSF and the misalignment of the slit relative to the parallactic angle at the high airmass of these observations (2.3 at epoch 1 and 1.8 at epoch 2). In the bottom panel the ratio between the two slit throughputs (solid line) is overplotted on the ratio between the spectra at the two epochs (dotted line).}
\label{fig1}
\end{figure}

\section{Observations and data reduction}

\subsection{\textit{Swift} detection}

The \textit{Swift} Burst Alert Telescope (BAT) triggered on \grb{} on February 10th, 2008 at $T_0=$ 07:50:06 UT \citep{Grupe08}. The duration spanning 90 per cent of the GRB emission (15--350 keV) was $45\pm11$ s and its 15--150 keV fluence was $1.8\times\,10^{-6}$ erg cm$^{-2}$. The time integrated BAT spectrum is best fit by a simple power law with photon index $\Gamma=1.77\pm0.12$ \citep{Ukwatta08}. An X-ray afterglow was observed with the \textit{Swift} X-Ray Telescope (XRT), starting 161 s (240 s) after the trigger in windowed timing (photon counting) mode. An optical afterglow was detected by the \textit{Swift} Ultra-Violet/Optical Telescope at a position R.A. $=16^\mathrm{h}45^\mathrm{m}04.01^\mathrm{s}$ and Decl. $=+13^\circ49\arcmin35.9\arcsec$ \citep[J2000, estimated 90 per cent confidence error radius of 0.6\arcsec;][]{Marshall08}. We retrieved the X-ray light curve and spectra from the \textit{Swift} repository \citep{Evans07,Evans09}.\\

\subsection{ULTRASPEC imaging}

\begin{figure*}
 \centering
 \includegraphics[width=180mm,angle=0]{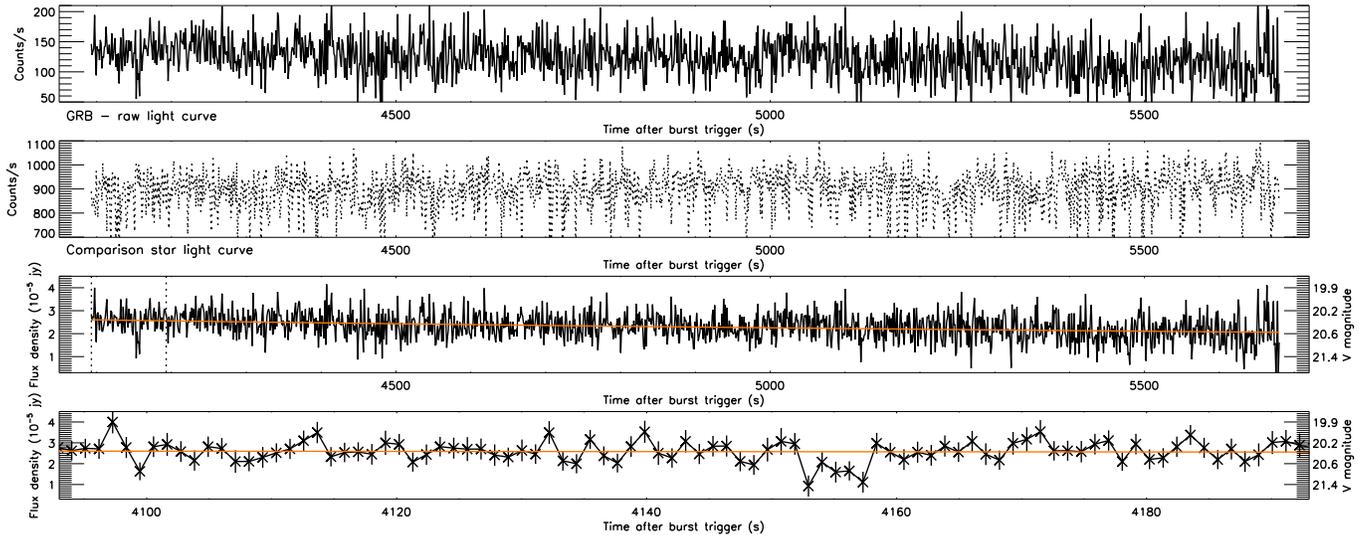}
\caption{The fast (1.09 s) sampled ULTRASPEC light curves ($V$ band). The whole observation (4093--5680 s after the trigger) is displayed in the three upper panels, while the first 100 s are shown in the bottom panel (1$\sigma$ errors over-plotted). The raw GRB (solid) and the comparison star (dotted) light curves are shown in the first and second panels, respectively. The GRB light curve, flux calibrated with the comparison star, is shown in panels 3 and 4. The flux decreases as a power law ($F \propto t^{-\alpha}$) with decay index $\alpha=0.74\pm0.07$ ($\chi^2_{\textrm{dof}}=1.03$ for 1453 degrees of freedom). Any short time-scale variation is consistent with statistical fluctuations smaller than $3\sigma$. }
\label{fig2}
\end{figure*}

For 26.45 min, starting at 08:58 UT on February 10th 2008, we observed \grb{} with ULTRASPEC at the ESO 3.6-m telescope on La Silla, Chile, mounted on the EFOSC2 spectrograph \citep{D'Odorico88}. Because of the frame-transfer capabilities of ULTRASPEC, it is possible to obtain very high time resolution data without sacrificing efficiency. Observations were taken in imaging mode, through the Bessel $V$-band filter. The CCD pixels were binned by $2\times2$, allowing $\sim$1 s sampling of the light curve, with negligible ($\sim 10$\,ms) dead time between exposures. Time-stamping of individual exposures uses a dedicated GPS-based system with a relative accuracy of 50 $\mu$s and an absolute accuracy of a few ms. The data were bias subtracted and subsequently flat-fielded using a median of 100 sky-flat frames.\\

The photometric information about the GRB afterglow was extracted using an implementation of the optimal photometry algorithm of \citet{Naylor98}, which provides significantly better signal-to-noise than aperture photometry for faint sources. A nearby comparison star was used both to estimate the point-spread function and to correct for transparency variations. The position of \grb{} was fixed with respect to the position of the comparison star; this ensures that centroiding on the faint GRB does not introduce spurious variability into the light curve. Observations of a flux standard were taken on the following night to place the measurements on a standard photometric system. We were not able to determine the $V$-band extinction coefficient for the night of these observations, so the La Silla average of 0.12 mag/airmass was used to correct for atmospheric extinction. The observation log is presented in Table \ref{tab1}.\\

\subsection{VLT/FORS2 observations}

Starting at 08:32 UT on 2008 February 10th  (42 min post-burst), a series of 600\,s spectra were obtained with VLT/FORS2 in long-slit spectroscopy mode with a 1\farcs0 wide slit, North-South oriented and centred with an $R$-band acquisition image. The sequence of grisms used was 300V, 600z+OG590, 1400V, 1200R+GG435 and finally 300V again. This allowed us to both cover a larger wavelength window with the lower resolution grism (300V) and obtain mid-resolution spectroscopy for different regions of the spectrum. The individual spectra were cleaned of cosmic rays using the Laplacian Cosmic Ray Identification algorithm of \citet{vanDokkum01}. The seeing remained relatively stable during the observations, between 1\farcs1 and 1\farcs4, yielding the spectral resolutions reported in Table \ref{tab1}. A first analysis of the spectrum revealed an absorption system associated with the host galaxy at redshift $z=2.641$ \citep{Jakobsson08a,Fynbo09}.\\

While the mid-resolution spectra were mainly used for the spectral analysis, we aimed to flux-calibrate the low-resolution 300V spectra for the SED study. However, the two 300V spectra were obtained at high airmass (2.3 and 1.8 respectively) where the difference between the slit position angle and the parallactic angle was 126.6$^\circ$ and 133.4$^\circ$, respectively. Thus, slit losses influence the continuum level of the spectra, particularly in the blue. In order to correct for this, we computed the slit throughput in the following way. A theoretical model of the point spread function (PSF) as delivered by an 8.2 m diameter Unit Telescope, including the central obscuration caused by the secondary mirror, was built using a piece of IDL code graciously made available by Enrico Fedrigo (private communication). In particular, this model includes the dependence of the diffraction-limited theoretical PSF as a function of wavelength. This model PSF was then convolved with a Gaussian whose wavelength-dependent FWHM follows Roddier's formula, i.e. $\propto (\lambda / \lambda_{\mathrm{ref}})^{-0.2}$, normalized to the value measured on each observed spectrum at the  effective wavelength of the $R$ filter used for the centering of the target ($\lambda_{\mathrm{ref}} =6600$ \AA{}).\\

For each spectrum, the distance of the PSF centre to the slit centre at a given wavelength was assumed to be the differential refraction between this wavelength and $\lambda_{\mathrm{ref}}$ multiplied by the cosine of the difference between the parallactic angle at the time of the observation and the parallactic angle at the time when the object would be at an airmass of 1.41, converted to degrees ($25^{\circ}$ and $18^{\circ}$). Here we assumed that the Longitudinal Atmospheric Dispersion Corrector (LADC) of FORS2 \citep*{Avila97} performs optimally up to airmass of 1.41.  For the calculation of the differential refraction, we assumed the usual atmospheric conditions at Paranal (temperature $ T = 12\,^{\circ}\mathrm{C}$, pressure $P = 743$ mbar). The integrated value of the flux along the spatial direction entering the spectrograph can then be calculated for each wavelength. The factor representing the slit throughput is then the ratio between this integrated value and the total flux at that wavelength. The top panel of Fig. \ref{fig1} shows the estimated slit throughputs as a function of wavelength for the two 300V FORS2 spectra. The bottom panel shows the ratio between the throughputs compared with the ratio between the two spectra. The agreement indicates that the slit throughputs have been reasonably well calculated. \\

The response curve correction was performed using observations of the standard star LTT3864. The flux calibration was rescaled using $R$-band VLT/FORS2 images. $R$-band photometry was secured with VLT/FORS2, before and after our spectroscopic observation. Calibration was carried out by observing the Landolt standard fields SA\,100 and Rubin\,149, which allowed us to obtain a photometric accuracy of 0.02 mag. The GRB observations were carried out at large airmass (1.7--2.5), but so was also one of the two standard fields, which allowed a reliable extinction coefficient to be computed, in agreement
with the value tabulated in the ESO web page. The photometric conditions were excellent according to the Paranal night logs.\\

\section{Data analysis}

\subsection{ULTRASPEC light curve}
\label{sec3p1}

\begin{figure}
 \centering
 \includegraphics[width=85mm,angle=0]{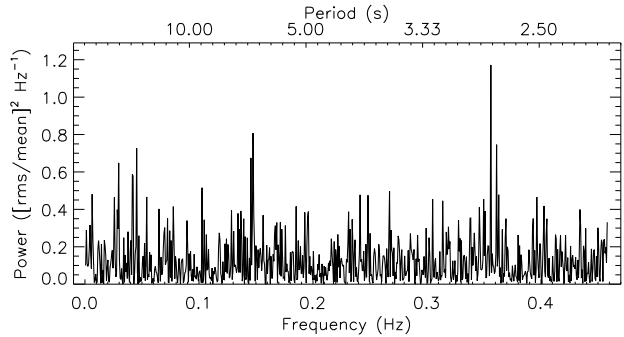}
\caption{The power spectrum of the ULTRASPEC light curve in fractional $\textrm{rms}^2$ units, after correcting for the decay. The minimum period detectable, given the time resolution, is 2.18 s, while the longest time scale monitored is 1587 s. The peaks in the power spectrum are generated by random noise, see Section \ref{sec3p1} for more details. The lack of a preferred frequency confirms the smoothness of the light curve.}
\label{fig3}
\end{figure}

The ULTRASPEC light curve is plotted in Fig. \ref{fig2}. We Fourier transformed the ULTRASPEC light curve in order to investigate the variability and possible periodicities over all time scales. We exclude the flux variation due to the afterglow natural evolution by first correcting the light curve with the power-law fit to the decay. Figure \ref{fig3} shows the power spectra of the unbinned light curve. Thanks to the fast sampling of ULTRASPEC, we can monitor the power spectrum down to 2.18 s time scales. The power spectrum of the light curve shows a peak at 2.8 s, likely due to random noise. Indeed, the peak height is not significant (3 per cent of the flux, corresponding to log $(\Delta F/F)= -1.57$, (log $\Delta t/t)=-3.2$ at 81.4 min), well below the instrument detection limit, see Sec. \ref{sec4p2}. We performed a Monte Carlo type generation of 100,000 light curves with random variability equal to the measured GRB variability and found peaks of this height and greater to occur in 6 per cent of cases. In addition, the phase-folded light curve does not show any evidence for periodicity. Thus, no frequency is preferred in the power spectrum. The ULTRASPEC light curve is smooth and follows a pure power law within the statistical fluctuations.\\

\begin{table*}
\begin{tabular}{ l | c l c c r}
\hline \hline
\rule[-0.2cm]{0mm}{0.6cm}
Line [$\lambda_{\textrm{vac}}$] & Observed wavelength & \multicolumn{1}{c}{Redshift} & Observed EW & Notes \\
\rule[-0.2cm]{0mm}{0.6cm}
(\AA{}) &  (\AA{}) &   & (\AA{}) & \\
\hline

\multicolumn{5}{c}{}\\
\multicolumn{5}{c}{1400V}\\
\multicolumn{5}{c}{}\\

\mbox{O\,{\sc i}}   $\lambda1302$   & 4740.6 & 2.6405  & $4.52\pm0.42$ &   \\
\mbox{Si\,{\sc ii}}  $\lambda1304$   & 4747.7 & 2.6398  & $4.81\pm0.41$ & \mbox{O\,{\sc i}*} $\lambda1304$ contribution \\
\mbox{C\,{\sc ii}}  $\lambda1334$   & 4858.4 & 2.6406  & $9.71\pm0.47$ &  \\
\mbox{C\,{\sc ii*}} $\lambda1335$  & 4864.1 & 2.6416  & $1.93\pm0.22$ &    \\
\mbox{Si\,{\sc iv}} $\lambda1393$  & 4890.5 & 2.5088$^a$  & $0.93\pm0.33$ & \mbox{Si\,{\sc iv}} $\lambda1402$ not detected\\
\mbox{Si\,{\sc iv}} $\lambda1393$  & 5073.8 & 2.6403  & $5.52\pm0.28$ &  \\
\mbox{Si\,{\sc iv}} $\lambda1402$  & 5106.7 & 2.6404  & $4.52\pm0.27$ & \\
\mbox{C\,{\sc iv}} $\lambda1548$  & 5431.8 & 2.5085$^a$  & $2.44\pm0.21$ &  \\
\mbox{C\,{\sc iv}} $\lambda1550$  & 5440.8 & 2.5084$^a$  & $1.93\pm0.23$ &   \\
\multirow{2}*{\mbox{Si\,{\sc ii}} $\lambda1526$} &5556.7 &  \multirow{2}*{\vline}\,2.6396  & \multirow{2}*{$5.53\pm0.34$} &  \multirow{2}*{Two velocity components}\\
 & 5558.7 &  \,2.6416  &  & \\
\mbox{C\,{\sc iv}} $\lambda1548$  & 5636.0 & 2.6403  & $9.56\pm0.30$ &   \\
\mbox{C\,{\sc iv}} $\lambda1550$  & 5644.9 & 2.6400  & $7.75\pm0.27$ &   \\
\multirow{2}*{\mbox{Fe\,{\sc ii}} $\lambda1608$} & 5854.4 & \multirow{2}*{\vline}\,2.6396  & \multirow{2}*{$3.54\pm0.38$} &  \multirow{2}*{Two velocity components}\\
 & 5857.8 & \,2.6416   & &  \\

\multicolumn{5}{c}{}\\
\multicolumn{5}{c}{1200R}\\
\multicolumn{5}{c}{}\\

\multirow{2}*{\mbox{Al\,{\sc ii}} $\lambda1670$} & 6081.0 & \multirow{2}*{\vline}\,2.6396  & \multirow{2}*{$6.08\pm0.45$} & \multirow{2}*{Two velocity components}\\
 & 6083.6 & \,2.6416   & &  \\
\multirow{2}*{\mbox{Si\,{\sc ii}} $\lambda1808$} & 6580.9 & \multirow{2}*{\vline}\,2.6396  & \multirow{2}*{$2.36\pm0.39$} &  \multirow{2}*{Two velocity components}\\
 & 6583.9 & \,2.6416 &  & \\
\multirow{2}*{\mbox{Al\,{\sc iii}} $\lambda1854$}& 6749.8 & \multirow{2}*{\vline}\,2.6396  & \multirow{2}*{$3.36\pm0.34$} &   \multirow{2}*{Two velocity components}\\
& 6753.9 & \,2.6416 & &  \\
\multirow{2}*{\mbox{Al\,{\sc iii}} $\lambda1862$}& 6780.0 & \multirow{2}*{\vline}\,2.6396  & \multirow{2}*{$1.37\pm0.35$} &  \multirow{2}*{Two velocity components}\\
& 6783.8 & \,2.6416 &   &  \\

\multicolumn{5}{c}{}\\
\multicolumn{5}{c}{600z}\\
\multicolumn{5}{c}{}\\
\multirow{2}*{\mbox{Fe\,{\sc ii}} $\lambda2344$} & 8532.1 & \multirow{2}*{\vline}\,2.6496  & \multirow{2}*{$5.35\pm0.30$} & \multirow{2}*{Two velocity components} \\
 & 8648.5 & \,2.6416  & &  \\
\multirow{2}*{\mbox{Fe\,{\sc ii}} $\lambda2374$} & 8642.8 & \multirow{2}*{\vline}\,2.6396  & \multirow{2}*{$3.96\pm0.38$} & \multirow{2}*{Two velocity components}\\
 & 8648.1 & \,2.6416  & &  \\
\multirow{2}*{\mbox{Fe\,{\sc ii}} $\lambda2382$} & 8671.5 & \multirow{2}*{\vline}\,2.6396  & \multirow{2}*{$7.23\pm0.41$} &  \multirow{2}*{Two velocity components}\\
 & 8675.9 & \,2.6416  & &  \\

\hline \hline
\end{tabular}
\caption{Absorption lines in the medium resolution 1400V, 1200R and 600z spectra. The redshifts for the two-component profiles (short vertical lines) are derived from a Voigt profile fit. Observer frame EWs with $1\sigma$ errors are reported. $^a$ Intervening system.}
\label{tab2}
\end{table*}

\begin{center}
\begin{table*}
\begin{tabular}{ l|c c c c}
\hline \hline
\rule[-0.2cm]{0mm}{0.6cm}
 Ion [transitions] &  Component a   &  Component b & Total column density & [X/H]\\ 
\rule[-0.2cm]{0mm}{0.6cm}
                           &    log ($N/$cm$^{-2}$)     &     log ($N/$cm$^{-2}$)   &     log ($N/$cm$^{-2}$)         &      \\    
\hline

\mbox{Al\,{\sc ii}} [1670]$^a$          &                             &                             & $>13.56$             & $>-2.79$ \\  
\mbox{Al\,{\sc iii}} [1854,1862]        & $13.77\pm0.08$ & $13.91\pm0.13$ & $14.14\pm0.08$  & $-2.21\pm0.13$ \\
\mbox{Fe\,{\sc ii}} [1608]                  & $15.72\pm0.37$ & $15.63\pm0.54$ & $15.98^{+0.37}_{-0.26}$  & $-1.42\pm0.33$ \\
\mbox{Si\,{\sc ii}} [1526$^b$, 1808] & $15.84\pm0.18$ & $15.96\pm0.16$ & $16.20^{+0.13}_{-0.11}$  & $-1.21\pm0.16$ \\
\mbox{Zn\,{\sc ii}} [2026]$^c$          &                             &                            & $13.53\pm0.14$  & $-0.93\pm0.18$ \\ 

\hline \hline 
\end{tabular}
\caption
{\protect The ionic column densities estimated from a simultaneous Voigt profile fit to the lines in the 1400V and 1200R medium resolution grism spectra. $^a$\mbox{Al\,{\sc ii}} $\lambda$1670 line is saturated, the lower limit on $N_{\textrm{Al\,{\sc ii}}}$ is derived from the EW (Table \ref{tab2}). $^b$The \mbox{Si\,{\sc ii}} $\lambda$1526 line is only included in a first stage to model the line profiles; the Si abundance is computed using only the weaker and non-saturated \mbox{Si\,{\sc ii}} $\lambda$1808 line. $^c$Zn abundance estimated from the low resolution 300V spectrum (b$_{\textrm{turb}}$=39.4, b$_{\textrm{th}}$=0 km s$^{-1}$).}
\label{tab3}
\end{table*}
\end{center}

\subsection{VLT/FORS2 spectral analysis}

\begin{figure}
 \centering
 \includegraphics[width=87mm,angle=0]{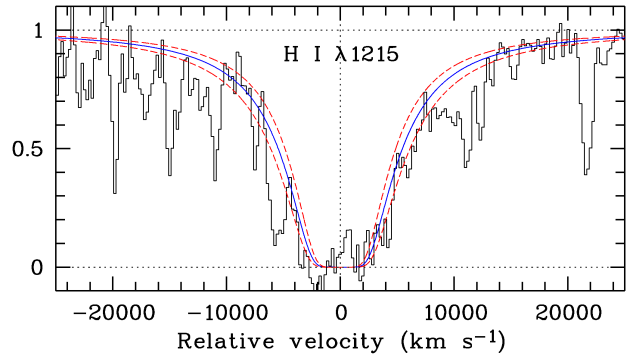}
 \caption{A portion of the normalized optical afterglow spectrum, centred on the Ly$\alpha$ absorption line, at the GRB host galaxy redshift. A neutral hydrogen column density fit to the damped Ly$\alpha$ line is shown with a solid line (log $(N_{\mathrm{H\,{\sc I}}}/$cm$^{-2})=21.90\pm0.10$), while the $1\sigma$ errors are shown with dashed lines.}
 \label{fig4}
\end{figure}

In the 300V combined spectrum (averaged from the two 300V flux calibrated spectra), we identify a damped Ly$\alpha$ absorber (DLA) in addition to a number of absorption lines at $z=2.641\pm0.001$, associated with the host galaxy of the burst. We also identify an intervening system at $z_{\textrm{int}}=2.509\pm0.001$ from \mbox{Si\,{\sc iv}} and \mbox{C\,{\sc iv}} transitions. A list of the lines detected in the low resolution 300V spectrum and their equivalent widths (EWs) is reported in \citet{Fynbo09}. We measured the EWs for both 300V epochs and find no evidence for spectral variability in the absorption. For the \mbox{H\,{\sc i}} DLA fit, we derived log $(N_{\mathrm{H\,{\sc I}}}/$cm$^{-2})=21.90\pm0.10$ (Fig. \ref{fig4}). The tentative detection of Ly$\alpha$ emission inside the DLA, as seen from the 1D spectrum \citep{Jakobsson08a}, is most likely noise, since it is not detected in the 2D frame.\\

A good metallicity estimator is usually the weak transition \mbox{Zn\,{\sc ii}} $\lambda$2026, but for \grb{} this line is only covered by the low resolution 300V spectrum, allowing only a crude metallicity estimate. From the simultaneous Voigt-profile fit of the \mbox{Zn\,{\sc ii}} $\lambda$2026 with the \mbox{Si\,{\sc ii}} $\lambda$1526, $\lambda$1808 and \mbox{Fe\,{\sc ii}} $\lambda$1608, $\lambda$1611 lines in the low resolution spectrum, performed with the MIDAS/FITLYMAN software \citep{Fontana95}, we derive a log $(N_{\mathrm{Zn\,{\sc II}}}/$cm$^{-2})=13.53\pm0.14$ cm$^2$ (i.e. [Zn/H] $-0.93\pm0.18$, Doppler thermal broadening $b_{\textrm{th}}=0$ km s$^{-1}$, and turbulent broadening $b_{\textrm{tur}}=39.4\pm6.8$ km s$^{-1}$). The analysis seems to show that the \mbox{Zn\,{\sc ii}} $\lambda$2026 line is on the linear part of the curve of growth (i.e. unsaturated), despite the low spectral resolution of the data. The metallicities refer to the solar abundances reported by \citet{Asplund09}.\\

\begin{figure}
 \includegraphics[width=85mm,angle=0]{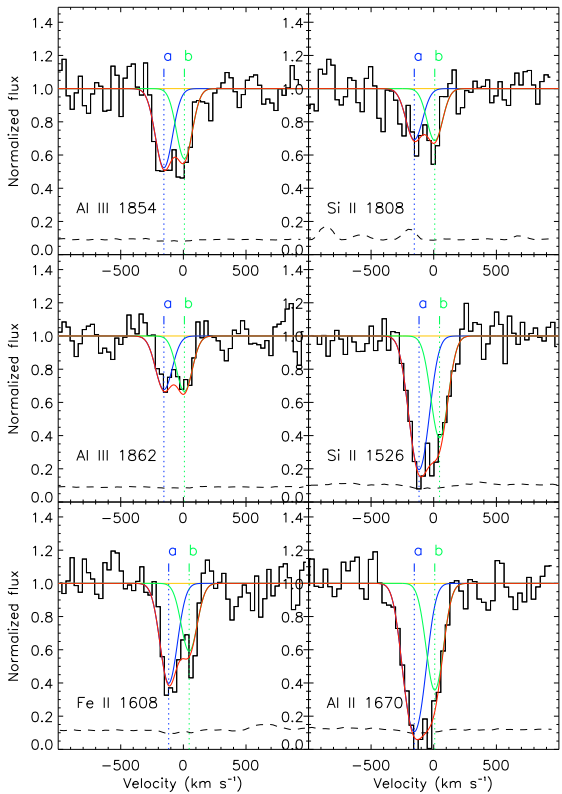}
\caption{The line profiles in the medium resolution 1400V and 1200R spectra are best modelled with two components separated by $148\pm25$ km s$^{-1}$ in velocity: $z_a=2.6396\pm0.0003$, $z_b=2.6416\pm0.0003$ (dotted lines), $b_{\textrm{turb, a}}=38\pm7$ km s$^{-1}$, $b_{\textrm{turb, b}}=23\pm6$ km s$^{-1}$, $b_{\textrm{th}}=0$ km s$^{-1}$. The dashed curves show the $1\sigma$ errors.}
 \label{fig5}
\end{figure}

Several lines are also identified in the medium resolution 1400V, 1200R and 600z  spectra with the EWs listed in Table \ref{tab2}. Many of the lines associated with the GRB host galaxy system show evidence for a two-component profile (component ``a" and ``b"). In order to derive reliable column densities and to study the kinematics of the gas, we select the \mbox{Fe\,{\sc ii}}, \mbox{Si\,{\sc ii}}, \mbox{Al\,{\sc ii}} and \mbox{Al\,{\sc iii}} lines in the higher resolution grisms 1400V and 1200R that are neither too saturated nor blended with other transitions, and model them simultaneously with a two-component Voigt profile, using the VPFIT\footnote{\texttt{Available at http://www.ast.cam.ac.uk/$\sim$rfc/vpfit.html}} software. In this way, the line profile of all the species is modelled with the same redshift $z$ and $b_{\textrm{tur}}$, for a given component, resulting in different column densities for different ions. We expect the \mbox{Fe\,{\sc ii}}, \mbox{Si\,{\sc ii}}, \mbox{Al\,{\sc ii}} and possibly \mbox{Al\,{\sc iii}} to be co-spatial and therefore to show a similar line profile. This is what we observe in the line-of-sight to \grb. The \mbox{Si\,{\sc ii}} $\lambda$1304 transition was excluded from the analysis because it is blended with the possibly dominating \mbox{O\,{\sc i}*} $\lambda$1304 line. The 600z spectrum was not included in the Voigt profile fit because of its poorer spectral resolution. The two components that model the line profiles are separated by $148\pm25$ km s$^{-1}$. The normalization was determined locally around each line and telluric features were excluded from the fit. Figure \ref{fig5} shows the two-component Voigt profile fit to the lines in the medium resolution spectra. The abundances and corresponding metallicities are presented in Table \ref{tab3}. Iron is probably depleted onto dust grains and is thus not a good metallicity indicator \citep{Savage96}. The best metallicity estimate is derived from silicon [Si/H] $=-1.21\pm0.16$ (1$\sigma$ uncertainties), corresponding to $Z/Z_{\odot}= 0.06^{+0.03}_{-0.02}$. \\

\subsection{Spectral energy distribution}
\label{sec3p3}

\begin{table*}
\begin{tabular}{ l | r r c c c c c }
\hline \hline
\rule[-0.2cm]{0mm}{0.6cm}

 Extinction &  Model &  \multicolumn{1}{c}{$\chi^2$ [dof]}  & $E(B-V)$  & \multicolumn{1}{c}{$N_\mathrm{H}$}  & $\beta_1$ & $\beta_2$ & $E_{\textrm{break}}$ \\

\rule[-0.2cm]{0mm}{0.6cm}
       type &        &                &  (mag) & ($10^{22}$ cm$^{-2}$)& &                &   (keV)      \\ 
\hline
\rule[-0.0cm]{0mm}{0.4cm}
 
           & PL                        &  517.2[39]  & $0.10\pm0.01$   & $<0.03$                            & $0.82\pm0.02$                 &       -                      &      -       \\
\rowcolor{yellow}SMC&BPL&  191.6[37]  & $0.06\pm0.01$   & $0.38^{+0.19a}_{-0.17}$   & $0.71\pm0.02$                 &  $1.59\pm0.11$       & $1.40^{+0.09}_{-0.13}$  \\
          & TBPL                     &  232.0[38]  & $0.07\pm0.01$   & $0.27^{+0.18}_{-0.17}$    & $\Gamma_2-0.5$               & $1.23\pm0.02$       &  $1.16^{+0.10}_{-0.12}$  \\
   & & & & & &   &   \\   

         & PL     &  434.9[39]  & $0.17\pm0.02$   & $<0.22$                           & $0.91\pm0.02$        &       -                              &      -       \\  
 LMC & BPL   &  194.4[37]  & $0.10\pm0.02$   & $0.48^{+0.19}_{-0.18}$    & $0.77\pm0.03$        & $1.59^{+0.11}_{-0.10}$ & $1.41^{+0.10}_{-0.12}$  \\
         & TBPL &  238.7[38]  & $0.10\pm0.02$   & $0$                                   & $\Gamma_2-0.5$     & $1.28\pm0.03$             &  $1.32\pm0.11$  \\
   & & & & & &   &   \\  

        & PL      &  546.9[39]  & $0.21\pm0.02$   & $<0.25$                            &  $0.91\pm0.03$              &       -                   &      -       \\
MW  & BPL     &  265.0[37]  & $<0.02$              & $0.20^{+0.18}_{-0.16}$    &  $0.63^{+0.03}_{-0.01}$  &  $1.59\pm0.11$  &  $1.39^{+0.09}_{-0.12}$ \\
        & TBPL   &  323.1[38]  &$0.05\pm0.03$   & $0$                                    & $\Gamma_2-0.5$                  &  $1.20\pm0.03$  &  $1.20\pm0.10$  \\
   & & & & & &   &   \\  

\hline \hline
\end{tabular}
\caption{Parameters resulting from a joint fit of the optical-to-X-ray SED at 66 min after the burst, assuming an absorbed PL, BPL or TBPL, see main text. The host galaxy dust extinction is modelled with a SMC, LMC or MW extinction law (Pei 1992) and the excess X-ray absorption is measured assuming solar metallicity. The best fitting model is highlighted. All the errors refer to a 90 per cent confidence level. $^a$ $N_\mathrm{H}=2.04^{+1.03}_{-0.95}\times 10^{22}$ cm$^{-2}$ for $Z/Z_\odot=0.06$.}
\label{tab4}
\end{table*}

\begin{figure*}
 \includegraphics[width=100mm,angle=270]{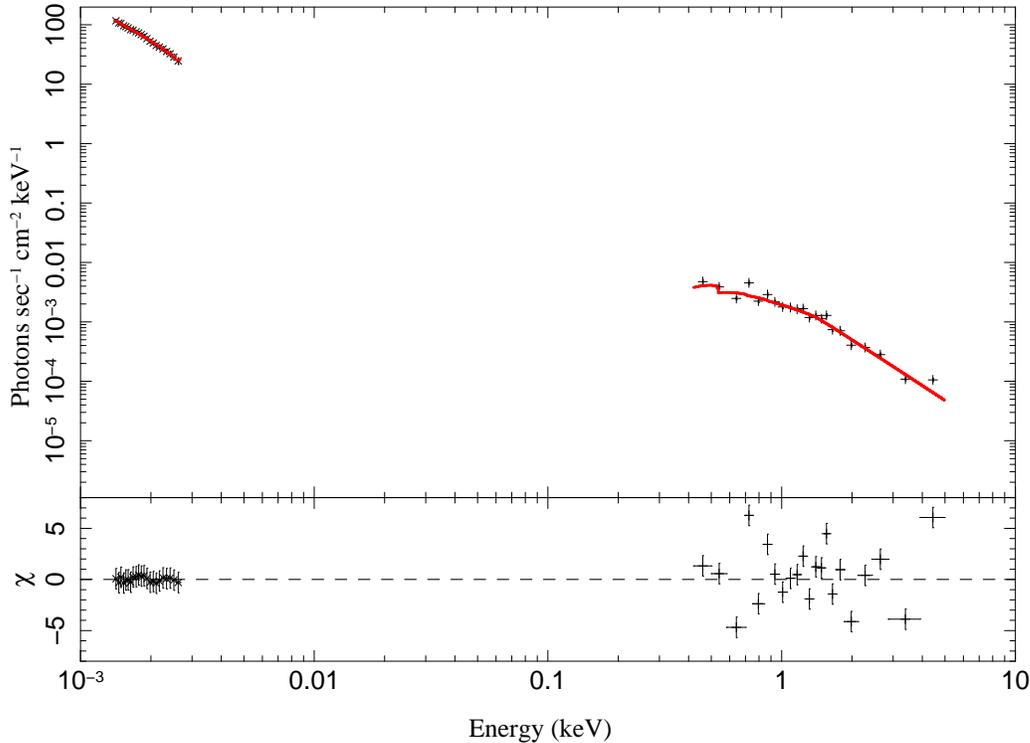} 
 \caption{\protect The SED of the optical 300V spectrum ($\times$) and the X-ray spectrum (+), at 66 min after the burst. The solid line shows the best fitting model, a broken power law with spectral slopes $\beta_{\mathrm{X}}=1.59\pm0.11$ and $\beta_{\mathrm{opt}}=0.71\pm0.02$, for an SMC-type dust extinction, see Table \ref{tab4}. The residuals are displayed in the bottom panel.}
 \label{fig6}
\end{figure*}

\begin{figure*}
\centering
 \includegraphics[width=178mm,angle=0]{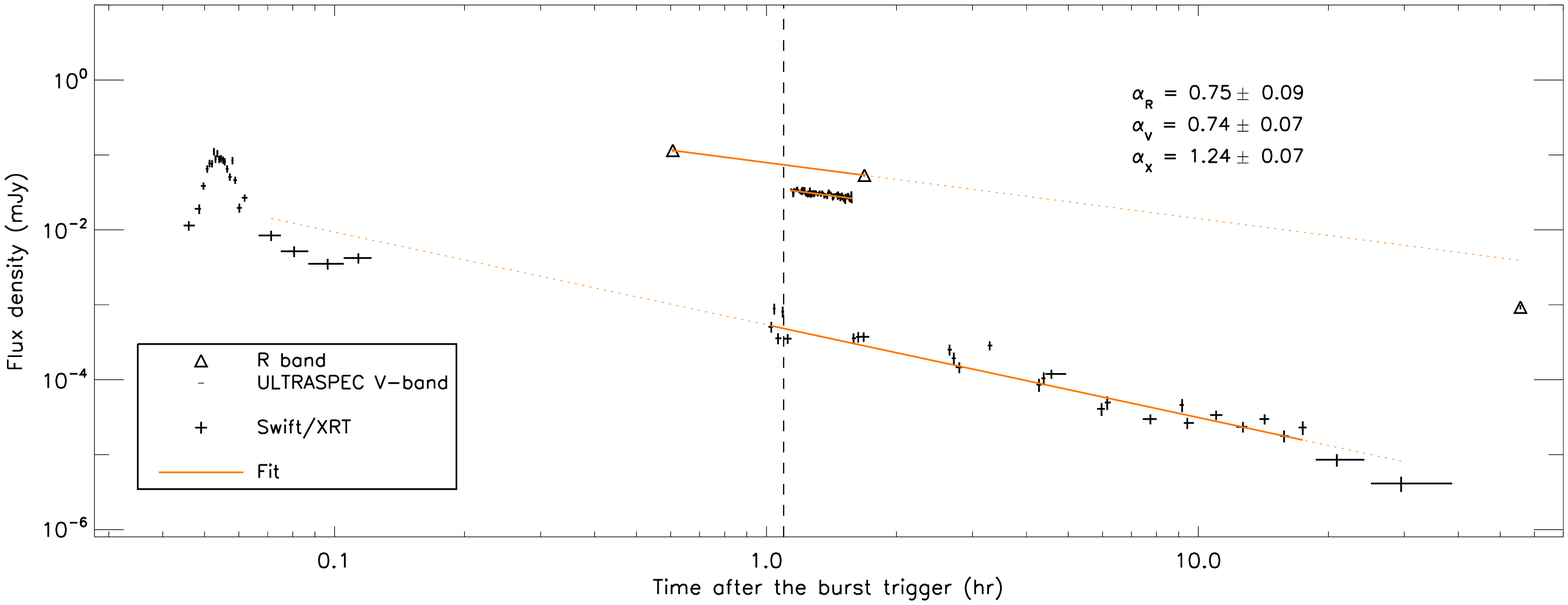}
\caption{The optical and X-ray afterglow light curves of \grb{}. The \textit{Swift}/XRT X-ray light curve (at 1.73 keV) is plotted beneath the optical data. The ULTRASPEC $V$-band light curve is here plotted with a bin factor of 10. The $R$-band decay is derived from the VLT/FORS2 data. Late time points were excluded to avoid a possible break. The solid lines show the fit to the data, while the dotted lines extrapolate the fit to the complete datasets. The vertical line shows the SED time. The errors are 1$\sigma$.}
\label{fig7}
\end{figure*}

In order to fit the optical-to-X-ray SED, we combine the averaged 300V optical with the X-ray spectra. We choose to include the 300V spectrum in the SED to investigate the significant dust reddening reported by \citet{Fynbo09}. Also, the larger spectral window coverage of the 300V as compared to the higher resolution spectra makes it more suitable to investigate the SED. The SED time is chosen at the logarithmic mean between the two 300V observations (3945 s). Using the \textit{Swift} spectrum repository \citep{Evans07,Evans09}, we extract the X-ray spectrum from a narrow time interval (3690--4200 s, logarithmically centred on the SED time), and use its count rate to scale the X-ray spectrum extracted from a larger time window (3690--106130 s) and with a better signal-to-noise (S/N). This approach assumes no spectral evolution in the X-ray spectrum, as confirmed by extracting spectra for different time windows and by the constant hardness ratio. This allows the optical and the X-ray spectra to be compared in flux. The 300V averaged spectrum was cleaned (absorption lines removed and frequencies bluer than Ly$\alpha$ excluded) and corrected for Galactic extinction. The optical spectrum was then binned into 22 bands (192 \AA{}) in order to obtain the same number of data points as in the X-ray spectrum. The statistical errors were calculated from the variance in the spectrum. A systematic error introduced by the response function was calculated by measuring the amplitude of the spurious wiggles introduced by the flux calibration process. This uncertainty and the error due to the calibration against the VLT photometric measurement were then added in quadrature to the formal error. The total resulting error on the optical spectrum is about 8 per cent of the flux. The optical-to-X-ray SED is shown in Fig. \ref{fig6}.\\

We model the SED from the optical to the X-rays with  a single power law (PL), a broken power law (BPL) and a tied broken power law (TBPL), where the spectral slopes are tied to differ by $\Delta\beta=0.5$ to reproduce the spectral break (cooling frequency) expected for a synchrotron spectrum. The fit was performed with the Interactive Spectral Interpretation System \citep[ISIS;][]{Houck00} software, which allows all the data to be compared directly in count space. Working in count space has the advantage of not requiring any a priori model for the X-rays, otherwise needed for the conversion of the X-rays into flux \citep[see e.g][]{Starling07}. The host galaxy dust extinction was modelled with Small Magellanic Cloud (SMC), Large Magellanic Cloud (LMC) or Milky Way (MW) extinction curves \citep{Pei92} and the X-ray absorption \citep[in excess of the Galactic $N_{\textrm{H,Gal}}=(5.59\pm0.02)\times10^{20}$ cm$^2$;][]{Kalberla05} is measured from a \emph{zphabs} model in ISIS, assuming Solar metallicity. Table \ref{tab4} summarizes the fit results.\\

The best fitting model is a BPL with spectral slopes $\beta_{\mathrm{X}} = 1.59\pm0.11$ and $\beta_{\mathrm{opt}} = 0.71\pm0.02$, with the cooling frequency occurring at $E_{\textrm{break}}=1.40^{+0.09}_{-0.13}$ keV in the soft X-ray range. These values are derived for SMC-like extinction (lowest $\chi^2=191.6$, for 37 degrees of freedom), while the LMC extinction curve provides a very similar fit, resulting in consistent parameter values. The MW extinction curve and its related 2175 \AA{} bump can be excluded. The optical spectrum is reddened by dust grains at the redshift $z=2.641$ of the host galaxy, with $E(B-V)=0.06 \pm 0.01$ mag, or $A_V=0.18\pm0.03$ mag (rest-frame), modelled with an SMC-like extinction curve.  We find an excess X-ray absorption of $N_\mathrm{H}=0.38^{+0.19}_{-0.17}\times 10^{22}$ cm$^{-2}$ assuming Solar metallicity, whereas $N_\mathrm{H}=2.04^{+1.03}_{-0.95}\times 10^{22}$ cm$^{-2}$ for $Z/Z_\odot=0.06$. The above errors refer to a 90 per cent confidence level. The SED results should be treated with caution, because they are dependent on the slit loss correction of the 300V spectra. In particular, the same SED analysis, but for the optical spectrum that has not been corrected for slit losses, provides a 4 per cent change in the optical slope and 68 per cent in the $E(B-V)$, for the best fit model.\\

\subsection{Afterglow evolution}
\label{sec3p4}

\begin{table}
\begin{center}
\begin{tabular}{ c c r}
\hline \hline
\rule[-0.25cm]{0mm}{0.6cm}

Time since GRB (hr) & Instrument    & \multicolumn{1}{c}{Magnitude} \\  

\hline

0.61   &   VLT/FORS2          &  18.74$\pm$0.05  \\

1.69  &     VLT/FORS2         &  19.57$\pm$0.05\\

55.70  &  Keck-I/LRIS$^a$    &  23.97$\pm$0.07\\

\hline \hline 
\end{tabular}
\caption
{\protect The $R$-band photometry, not corrected for Galactic extinction (1$\sigma$ errors). $^a$ Perley \& Bloom, private communication.}
\label{tab5}
\end{center}
\end{table}

Figure \ref{fig7} shows the afterglow time evolution in the X-ray and optical bands. We converted the X-ray light curve into monochromatic flux at 1.73 keV, the logarithmic average of the XRT band, assuming a spectral slope $\beta_\mathrm{X}=1.59$ as derived from the SED (see Section \ref{sec3p3}). Early ($<12$ min) X-ray data were excluded from the fit to avoid the influence of flares. The last XRT data points were also excluded in order to avoid the contribution from a possible break at late time that cannot be constrained. We fit the X-ray light curve with a single power law with temporal slope $\alpha_\mathrm{X}=1.24\pm0.07$ (reduced $\chi^2_\nu=3.67$ for 23 dof), noting that a broken power law does not improve the fit. The high $\chi^2_\nu$ could be produced by the wiggles observed in the X-ray light curve, possibly originated by micro-variability. However, the poor sampling of the X-ray light curve does not allow us to investigate this further.\\

The $V$-band temporal decay was derived from a power-law fit to the ULTRASPEC light curve $\alpha_{V}=0.74\pm0.07$, see Section \ref{sec3p1}. We collected the $R$-band photometric data points from our VLT/FORS2 acquisition images and the Keck-I/LRIS data, reported in Table \ref{tab5}, corrected them for Galactic extinction \citep*[$A_V = 0.276$ mag;][]{Schlegel98}, and converted them to flux density. A temporal decay with slope $\alpha_{\textrm{VLT+Keck}}=1.07\pm0.07$ can be derived by a poor power-law fit to the three $R$-band data points (reduced $\chi^2_\nu=26.5$), in disagreement with the $V$-band decay. This suggest the presence of a break in the light curves at late times. Thus, the Keck data point was excluded from the temporal decay study to avoid the contribution from the possible break.  The $R$-band decay derived from the two VLT data points has a temporal slope $\alpha_{R}=0.75\pm0.09$, consistent with the $V$ band, where the error was calculated from the minimum and maximum slopes between the two points.\\

\section{Discussion}

\subsection{Modelling the afterglow}
\begin{table*}
\begin{center}
 \begin{tabular}{ l | c c c c c c c c}
\hline\hline
\rule[-0.25cm]{0mm}{0.6cm}
 & $\alpha_{\textrm{obs}}$ & $\beta_{\textrm{obs}}$ & Regime & $\alpha(\beta)_{\textrm{exp}}$ & $\beta(\alpha)_{\textrm{exp}}$ & $p(\alpha)$& $p(\beta)$& $\sigma_{p(\alpha),p(\beta)}$\\ 
\hline
 \rule[-0.4cm]{0mm}{1.cm} 
Optical & $0.75\pm0.09$ & $0.71\pm0.01$  & $\nu<\nu_c$  & $1.07\pm0.02$ & $0.50\pm0.06$  & $2.00\pm0.12$ &  $2.43\pm0.02 $& 3.5 \\
\rule[-0.4cm]{0mm}{1.cm} 
X-rays& $1.24\pm0.07$ & $1.59\pm0.07$ & $\nu_c<\nu$  & $1.88\pm0.11$ & $1.16\pm0.05$  & $2.32\pm0.09$ & $3.18\pm 0.14$ & 5.2 \\
\hline\hline
\end{tabular} 
\caption{The optical and X-ray temporal and spectral indices $\alpha$ and $\beta$ as observed and expected from the fireball model. We assume here an ISM scenario, with no extra energy injection and the slow cooling regime \protect\citep[e.g.,][]{Zhang06}. The errors are $1\sigma$. The electron energy distribution indices $p(\alpha)$, derived from the temporal slope, agree within $2.1\sigma$ between optical and X-rays, while $p(\beta)$ disagree at a $5.3\sigma$ level. The level of agreement between $p(\alpha)$ and $p(\beta)$ is indicated by $\sigma_{p(\alpha),p(\beta)}$.}
\label{tab6}
\end{center}
\end{table*}

In order to investigate the physics of the \grb{} afterglow, we attempt to model it within the synchrotron scenario. In Table \ref{tab6} the temporal slope $\alpha$ and the spectral slope $\beta$ are collected from both the optical and X-ray analysis (where $F_{\nu}\,{\propto}\,t^{-\alpha}\,\nu^{-\beta}$), as derived above. We first note that the $\beta_\mathrm{X} - \beta_{\mathrm{opt}}=0.88\pm0.07$ disagrees with the $\Delta \beta=0.5$ expected from the fireball model. In particular, this implies that the spectral break is not a cooling break and that the optical and the X-ray emission are not produced by a coherent synchrotron process. Possibly, the optical radiation and the X-rays were emitted in different regions, the overall SED resulting from a composition of two synchrotron spectra. Alternatively, different radiative processes must be invoked to explain the SED.\\ 

We further test fireball model predictions calculating the electron energy distribution index, $p$, from the temporal and spectral indices, see Table \ref{tab6}. We assuming a simple ISM, slow cooling scenario, with no extra energy injection \citep{Zhang06} and the cooling frequency in the soft X-rays, as derived by the SED fit. The electron indices derived from the temporal and spectral slope show a poor agreement ($3.5\sigma$) for the optical band,  and disagreement ($5.2\sigma$) for the X-rays. Although the optical and X-ray temporal slopes provide a similar $p(\alpha)$ (within $2.1\sigma$), the electron indices derived from the spectral slopes disagree at a $5.3\sigma$ level between the optical and the X-rays.\\ 

Thus, the closure relations are not satisfied for the case of \grb{}. In particular, the X-ray spectral slope seems too steep to be produced by the synchrotron electron cooling expected in the model. One possible reason for this is that the X-ray spectral slope was overestimated due to the degeneracy with the spectral break and the X-ray absorption. However, using the optical data in the SED helps in breaking this degeneracy. This suggests that the the fireball model cannot properly reproduce the \grb{} afterglow and therefore cannot be applied to the data. An independent SED study of a larger sample of GRB afterglows shows similar results for \grb{} (Zafar et al., in preparation).\\ 

\subsection{Variability}
\label{sec4p2}

The ULTRASPEC capability of observing at 1 s time resolution is a new frontier in the GRB afterglow variability study. But do we expect to see variability on those short time scales? How strong? And what processes can produce such variations? Answering these questions is essential to interpret not only the current light curve but also future observations with ULTRASPEC or any equivalent instrument. In order to address these questions and investigate the ULTRASPEC possibility of detecting fast variability, we analyse here the variability limits, derived by \citet{Ioka05}, based on kinematic arguments, showing that only certain time-scale fluctuations are physically allowed, at a particular observing time. These authors consider: (a) dips in the light curve, (b) bumps produced by density fluctuations, (c) a patchy-shell and (d) a refreshed shock. For the sake of clarity, we report below the limits from \citet{Ioka05} that are relevant for this paper.\\ 
 
 (a) The fluctuations that could produce dips in the light curve are limited to 
\begin{equation*}
\frac{\vert \Delta F_\nu \vert}{F_\nu}\leq\frac{4}{5}\left( \frac{\Delta t}{t}\right) ^2
\end{equation*} 
as derived from geometric constraints on the evolving emitting surface, considering causality arguments, relativistic effects and assuming a sudden shut off of the emission to obtain the upper limit on the variability.\\

(b) Regardless of their properties, the density enhancements can decelerate the emitting matter, limiting the variability to
\begin{equation*}
\frac{\vert \Delta F_\nu \vert}{F_\nu}\leq\frac{8}{5}\frac{\Delta t}{t}
\end{equation*} 
assuming the same geometric and causality arguments as above, and that the kinetic energy $E_{\textrm{kin}}$ is uniformly distributed in the variable volume.\\

(c) In case of a patchy shell, the time scale of the fluctuations is initially constrained to grow linearly in time \citep[$\Delta t\sim t$][]{Nakar04}, limiting the variability time scales to 
\begin{equation*}
\frac{\Delta t}{t}\geq 1
\end{equation*} 
for persistent angular fluctuations. \\

(d) Refreshed shocks can produce bumps with time scales 
\begin{equation*}
\frac{\Delta t}{t}\geq \frac{1}{4}
\end{equation*}
 if the acceleration of the GRB ejecta is hydrodynamic, as a slow shell will expand with its co-moving sound speed and collide with the decelerating leading shock-front.\\

If the emitting region is observed off-axis, i.e. when the line of sight is not aligned with the jet axis, and many regions ($>10^3$) contribute to the variability, the dips and density fluctuations, respectively, are limited to\\

(a*)
\begin{equation*}
\frac{\vert \Delta F_\nu \vert}{F_\nu}\leq\frac{6}{\sqrt{2}}\left( \frac{\Delta t}{t}\right) ^{3/2}
\end{equation*} 

(b*)
\begin{equation*}
\frac{\vert \Delta F_\nu \vert}{F_\nu}\leq 24\frac{\Delta t}{t}
\end{equation*} 
as derived by \citet{Ioka05} from cases (a) and (b) above.\\ 
 
\begin{figure}
 \centering
 \includegraphics[width=85mm,angle=0]{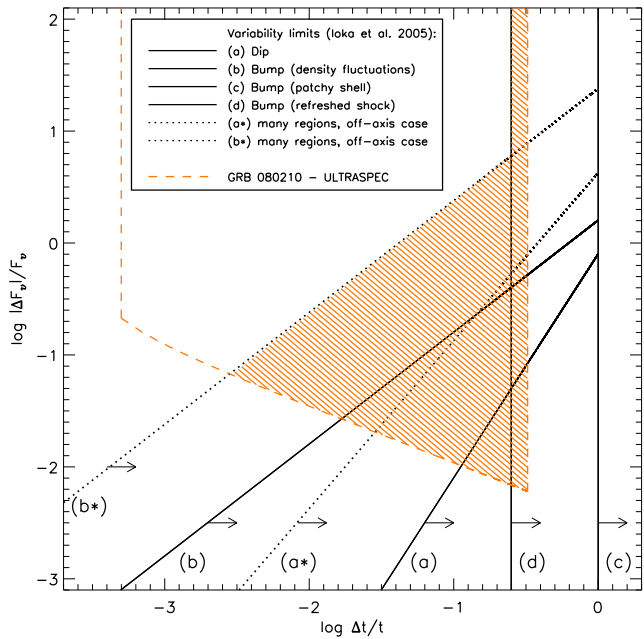}
\caption{Adapted from Ioka et al. (2005). The axes show the relative flux variation amplitude, $\vert\Delta F_\nu\vert/F_\nu$, versus the variability time scales over the time of observation, $\Delta t/t$. The solid lines reflect the variability limits derived from dips in the light curves (a), for bumps produced by density fluctuations (b), a patchy shell (c) and a refreshed shock (d). The dotted lines refer to the case of many fluctuation regions and off-axis observations, for dips (a*) and for density fluctuations (b*). The regions where variability is allowed by each process are indicated by the arrows. The \grb{} ULTRASPEC observation time-scale domain ($2.18$ s $< \Delta t < 26.45$ min at mid-exposure time after the burst, $t=81.44$ min, in the observer frame), is enclosed by the dashed lines. The variability region that is both physically allowed and detectable by ULTRASPEC is highlighted.}
\label{fig8}
\end{figure}
 
The variability limits discussed above are plotted in Fig. \ref{fig8} \citep[adapted from][]{Ioka05},  where the regions of allowed variability are indicated by the arrows, for each process. We also indicate the ULTRASPEC observation time domain (outlined by the dashed lines) at mid-exposure time ($81.44$ min), considering the covered time scales above $2.18$ s ($0.03~{\textrm{ min}}<\Delta t<26.45$ min) and the instrument detection limit. This limit is calculated from the light curve S/N over a single data point (time bin unit $dt=1.09$ s) and extended to longer time scales ($\Delta F / F ({n_{\textrm{bins}}}) = \Delta F / F (\textrm{1 bin}) / \sqrt{n_{\textrm{bins}}}$), where $n_{\textrm{bins}}$ is the number of time bins for each time scale. It is this detection limit that defines which fluctuations could possibly have been detected in the ULTRASPEC observations. The region where the allowed variability overlaps with the ULTRASPEC monitoring  is highlighted in Fig. \ref{fig8}. Given the smoothness of the \grb{} light curve, we can exclude any variability in this region, as it is physically allowed but not detected by ULTRASPEC.\\

A number of remarks can be deduced from Fig. \ref{fig8}. The fastest variability, both allowed and observable, can be produced by many density fluctuation regions (b*, upper dotted line in Fig. \ref{fig8}). For a single density fluctuation region (b, solid line), the S/N of this ULTRASPEC observation can probe variability only on time scales $\Delta t>72.8$ s. While these limits can provide constraints on the fluctuation amplitudes to be expected in a standard afterglow, they cannot easily be applied to \grb, as this afterglow does not seem to fit the synchrotron model. Nevertheless, the limits on dips in the light curve (a and a* lines) do not depend strictly on the fireball model, they only assume a relativistically expanding shell, regardless of the particular model \citep*{Fenimore96}. Thus, given the smoothness of the \grb{} light curve, we can limit the possible dips in the light curve to be weaker than 1 per cent in flux, on time scales longer than $\Delta t>9.2$ min. In case of many regions contributing to the dips in the light curve, we can exclude fluctuations stronger than 2 per cent of the flux on time scales $\Delta t>2.3$ min. These constraints are derived from the intersection between the ULTRASPEC detectability limit and the theoretical limits on light curve dips (a and a* lines). These limits on the variation amplitude can be interpreted in terms of under-density of the circumburst region within the fireball model. However, we cannot apply this to the case of \grb{} because of its non-standard afterglow physics. Finally, with the current ULTRASPEC dataset, refreshed shocks could in principle have been detected on time scales $20.00$ min $<\Delta t<26.45$ min, but they were not observed.\\ 

\subsection{Host galaxy environment}

\subsubsection{Gas location, metallicity and dust}
\label{sec4p3p1}

The spectroscopy of the optical afterglow reveals a number of absorption lines due to neutral and low-ionization species, i.e. \mbox{O\,{\sc i}}, \mbox{Si\,{\sc ii}}, \mbox{C\,{\sc ii}}, \mbox{Fe\,{\sc ii}}, \mbox{Al\,{\sc iii}} and \mbox{Zn\,{\sc ii}}, which can be used to investigate the properties of the absorbing region. The ionization potential of \mbox{O\,{\sc i}} (13.618 eV) is very close to that of \mbox{H\,{\sc i}} (13.598 keV): this already suggests that the two species could be co-spatial. Indeed, in low ionization media \mbox{O\,{\sc i}} and \mbox{H\,{\sc i}} tend to couple due to charge exchange \citep{Field71}. The ionization potentials of neutral Si, C, Fe, Al and Zn are well below 13.618 eV. Thus, all the observed species may, in principle, coexist in the same region. However, since the oxygen and hydrogen lines that we detect in the spectrum are saturated, their profiles cannot be used to compare the kinematics. On the other hand, the Voigt profile fit to the \mbox{Fe\,{\sc ii}}, \mbox{Si\,{\sc ii}} and \mbox{Al\,{\sc ii}} transitions indicates that these ions share the same two-component profile. Furthermore, they have comparable ionization potentials. These two pieces of evidence strongly suggest that these species are co-spatial. \mbox{Al\,{\sc iii}} is mildly ionized, and therefore belonging to a different gas-phase; however, its double velocity profile indicates that \mbox{Al\,{\sc iii}} is still related to the rest of the gas, possibly surrounding the bulk of the \mbox{H\,{\sc i}}.\\

Regarding the distance of the burst to the absorber, to first order, we can exclude that the lines are produced in the close vicinity of the GRB, as ionization is expected to occur inside $\sim$10 pc \citep[see eg.][]{Ledoux09}. The \mbox{Mg\,{\sc i}} is a possible distance limit indicator \citep*[$>50$ pc;][]{Prochaska06}, but none of its transitions were covered by the observations. The actual distance of the bulk of the gas may be much larger. Indeed, absorption systems have been found up to several kpc from the burst \citep{Vreeswijk07,D'Elia09,Ledoux09}, where the distance was computed using a photo-excitation (UV pumping) model of the fine-structure line variability. We also detect fine structure lines \citep[i.e. \mbox{Si\,{\sc ii}}*, \mbox{C\,{\sc ii}}* and \mbox{Fe\,{\sc ii}}* in the 1400V and 300V grisms, see also][]{Fynbo09}, but the low resolution of the FORS spectra does not allow any further modelling.\\

From the DLA profile fit, we derived a neutral hydrogen column density, log $(N_{\mathrm{H\,{\sc I}}}/$cm$^{-2}) =21.90\pm0.10$. This fairly high column density, compared to the low-resolution afterglow sample analysed by \citet{Fynbo09}, causes the neutral hydrogen to screen heavier elements (present in the same gas with much lower abundances)  from ionization. Thus, we assume that no ionization effects can significantly influence the metallicity estimate. The best metallicity indicator between the optical absorption lines that we detected is \mbox{Si\,{\sc ii}} $\lambda$1808, for which we find [Si/H$]=-1.21\pm0.16$ ($Z/Z_{\odot}= 0.06^{+0.03}_{-0.02}$). This suggests a chemically poor environment, quite common for GRBs with bright optical afterglows, where metallicities fall below $0.3Z_{\odot}$ for most absorbers \citep{Fynbo06}. \\

From the X-rays, we derive an equivalent hydrogen column density of log $(N_{\mathrm{H}}/$cm$^{-2})=21.58^{+0.18}_{-0.26}$, assuming Solar abundances, whilst log $(N_{\mathrm{H}}/$cm$^{-2})=22.31^{+0.18}_{-0.27}$ for $Z/Z_{\odot}=0.06$. The soft X-ray absorption is normally produced by metals in the line-of-sight, e.g., carbon and oxygen \citep{Wilms00}. The equivalent hydrogen column density measured from the X-ray absorption in \grb{} is comparable to the neutral hydrogen column density log $(N_{\mathrm{H}}/$cm$^{-2})=21.90\pm0.10$ from the Ly$\alpha$. However, in general the equivalent and neutral hydrogen column densities correlate extremely poorly \citep{Watson07}.\\ 

We find a visual extinction, $A_V=0.18\pm0.03$ mag, from the SED fitting and interpret it as due to dust. This value is quite common in GRB afterglows (\citealt{Kann10,Schady10}; Zafar et~al., in preparation.). An SMC extinction law  best reproduces the dust extinction that affects the \grb{} afterglow spectrum and a MW extinction law can be excluded. Consistent with this, we do not observe the 2175 \AA{} bump (7919 \AA{} in the observer frame), which is a typical signature of the Galactic dust absorption \citep{Pei92}. Even though such a structure has been observed in GRB afterglows \citep{Kruhler08,Eliasdottir09,Perley10}, it has been shown that in most GRB host galaxies, dust typically displays an SMC extinction curve \citep[e.g.,][]{Starling07,Kann10}. While a Galactic extinction law requires roughly the same amount of graphite and silicate grains, the SMC curve can be produced by silicate grains alone \citep{Pei92}. Thus, we infer a low graphite dust content for the \grb{} host galaxy. Although the presence of dust is expected in DLAs \citep{Pettini97}, the low metallicity disfavours the production of dust grains, as shown by the relation between dust and metallicity \citep{Vladilo98}.\\ 

\subsection{Origin of the intervening system}

The intervening system at $z=2.508$ would require a relative velocity $ v\sim11,000$ km s$^{-1}$, if associated with the GRB host galaxy. Velocities up to $3,000$ km s$^{-1}$ have been observed in GRB afterglow spectra \citep[e.g.,][]{Mirabal03} or expected by Wolf Rayet wind models \citep{VanMarle05}. However, none of the proposed scenarios seems to be able to reproduce $\sim10,000$ km s$^{-1}$.\\

 An intriguing possibility is that such an outflow could be accelerated by an active galactic nucleus (AGN). About half of the \mbox{C\,{\sc iv}} and \mbox{Mg\,{\sc ii}} narrow-line absorbers towards QSOs with apparent outflow velocities of 3,000--12,000 km s$^{-1}$ are actually intrinsic to the QSO/host \citep{Wild08}. We cannot exclude the possibility that the host galaxy of \grb{} is a low-luminosity AGN, since its emission would fall well below the detectability limit of X-ray telescopes. However, it is unlikely that a $\sim10,000$ km s$^{-1}$ fast outflow could remain as narrow in velocity as we observe ($b_{\textrm{turb}}\sim30$ km s$^{-1}$). In addition, if the faster outflows occur in the polar direction of an axisymmetric accretion geometry, it might be difficult to locate a star forming region hosting the burst between the nucleus and the accelerated absorbing material without invoking a fine tuned geometry. Moreover, the fastest outflows are typically located very close to the AGN itself and the burst location is unlikely to cross them on the line of sight.\\

Thus, the most favoured origin of the intervening system at $z=2.509$ is an absorber on the line of sight, a cloud or a galaxy $\sim43$ Mpc from the GRB host galaxy and unrelated to it. Statistically, a significant incidence of intervening systems is expected. Given the number density per unit redshift interval of intervening absorbers, not associated with the host galaxy, the probability of detecting at least one random \mbox{C\,{\sc iv}} absorber of rest-frame EW$(\lambda1548)>0.4$ \AA{} is 34 per cent \citep{Chen07}.\\

\section{Summary}

We searched for short-term variability, down to 2.18 s, in the ESO 3.6-m/ULTRASPEC observations of the \grb{} optical afterglow. The light curve decays as a power law ($\alpha=0.74\pm0.07$) and appears smooth on all time scales. Nevertheless, the time-monitoring allows us to investigate the circumburst environment and the blast-wave propagation. Comparing our observation with the variability limits derived by \citet{Ioka05}, we can exclude dips in the light curve with amplitude stronger than 1 per cent of the flux on time scales $\Delta t>9.2$ min and stronger than 2 per cent on time scales $\Delta t>2.3$ min, for a single or many under-dense regions respectively.\\

The \grb{} optical and X-ray late afterglows decay with temporal slopes $\alpha_{\mathrm{opt}}=0.75\pm0.09$ and $\alpha_\mathrm{X}=1.24\pm0.07$. The spectral slopes $\beta_{\mathrm{opt}}=0.71\pm0.01$ and $\beta_{X}=1.59\pm0.07$ are derived from the joint optical-to-X-ray SED fit with a broken power law ($1\sigma$ errors). We evaluate these observations with the theoretical expectation of the standard model and find no agreement within $5.3\sigma$, suggesting that the \grb{} afterglow cannot be produced with the fireball model physics.\\

From the SED analysis, we find that the spectral break is located in the soft X-ray at $E_{\textrm{break}}=1.40^{+0.09}_{-0.13}$ keV, while the X-rays absorption indicates an excess equivalent hydrogen absorption of log $(N_{\mathrm{H}}/$cm$^{-2})=21.58^{+0.18}_{-0.26}$ assuming Solar abundances, and log $(N_{\mathrm{H}}/$cm$^{-2})=22.31^{+0.18}_{-0.27}$ for $Z/Z_{\odot}=0.06$ (90 per cent confidence level errors). Optical reddening ($A_V=0.18\pm0.03$ mag) is induced by SMC-like dust (low graphite content).\\

In the optical VLT/FORS2 spectra, we detect several metal absorption lines associated with the GRB host galaxy ($z=2.641$), as well as a DLA (log $(N_{\mathrm{H\,{\sc I}}}/$cm$^{-2})=21.90\pm0.10$). We find [Si/H] $=-1.21\pm0.16$ ($Z/Z_{\odot}= 0.06^{+0.03}_{-0.02}$) suggesting a low metallicity environment. A Voigt-profile fit of the medium resolution lines reveals a two-component profile, separated by $148\pm25$ km s$^{-1}$, possibly associated with two major clouds along the line of sight within the host galaxy.\\

\grb{} represents one of the first attempts to study fast variability in GRB afterglows. Although this particular case must be treated with caution, due to its non-standard afterglow physics, our analysis demonstrated that the expected short-term can be detected by using the high speed read-out of the ULTRASPEC camera, specially for bright afterglows with higher S/N.\\

\section*{Acknowledgements}
ADC acknowledges support from the University of Iceland Research Fund. PJ acknowledge support by a Marie Curie European Re-integration Grant within the 7th European Community Framework Program and a Grant of Excellence from the Icelandic Research Fund. The Dark Cosmology Centre is funded by the Danish National Research Foundation. VSD, SPL and TRM are supported by STFC. ULTRASPEC was funded by the EU-OPTICON programme. This work made use of data supplied by the UK \textit{Swift} Science Data Centre at the University of Leicester. We thank \'Ard\'is El\'iasd\'ottir, Gudlaugur J\'ohannesson, Evert Rol, Rhaana Starling and Simon Vaughan for helpful discussions, Daniel Perley and Josh Bloom for providing the Keck photometric datapoint, and the referee Bruce Gendre for a careful and constructive report, which significantly improved the paper.\\



\label{lastpage}
\end{document}